\documentclass[preprint2]{proto}
\usepackage{times}
\usepackage{graphicx}

\usepackage{natbib}
\bibpunct{(}{)}{;}{a}{,}{,}
\begin{document}

\title{\textbf{\LARGE Extreme Deuteration and Hot Corinos: the
  Earliest Chemical Signatures of Low-Mass Star Formation}}

\author {\textbf{\large Cecilia Ceccarelli}} 
\affil{\small\em Laboratoire d'Astrophysique de Grenoble}
\author {\textbf{\large Paola Caselli}} 
\affil{\small\em INAF Oss. Astrofisico di Arcetri}
\author {\textbf{\large Eric Herbst}} 
\affil{\small\em Dept. of Physics, The Ohio State University}
\author {\textbf{\large Alexander G. G. M. Tielens}} 
\affil{\small\em Space Research Organization of the Netherlands}
\author {\textbf{\large Emmanuel Caux}} 
\affil{\small\em Centre d'Etude Spatiale des Rayonnements}

\begin{abstract}
%\begin{list}{ } {\rightmargin 1in}
%{\leftmargin 0in}
\baselineskip = 11pt
\leftskip = 0.65in
\rightskip = 0.65in
%rule{4.75in}{0.5pt}
%\vskip 1pt
\parindent=1pc {\small Low-mass protostars form from condensations
inside molecular clouds when gravity overwhelms thermal and magnetic
supporting forces.  The first phases of the formation of a solar-type
star are characterized by dramatic changes not only in the physical
structure but also in the chemical composition.  Since PPIV
(e.g., {\it Langer et al.}), exciting new developments have occurred in our
understanding of the processes driving this chemical evolution.  These
developments include two new discoveries : 1) extremely enhanced
molecular deuteration, which is caused by the freeze-out of
heavy-element-bearing molecules onto  grain mantles during the Prestellar
Core and Class 0 source phases; and 2) hot corinos, which are warm and
dense regions at the center of Class 0 source envelopes and which are
characterized by a multitude of complex organic molecules.  In this
chapter we will review these two new topics, and will show how they
contribute to our understanding of the first phases of solar-type
stars.  \\~\\~\\~}
%leave this in to get the correct vertical space after the abstract

%\end{list}
\end{abstract}

\section{INTRODUCTION}\label{sec:introduction}

Molecular clouds are the placentas inside which matter evolves from
embryos to stars and planetary systems. The material in the molecular
cloud feeds the newly forming system.  During this formation, the material
undergoes several dramatic changes. This chapter focuses on the
changes of the chemical composition during the first phases of star
formation.

There are several reasons why the study of chemistry in the first
phases of star formation is fascinating and important. Among them,
two stand out. First, {\it chemistry is a very powerful diagnostic},
both of the current and the past physical conditions of the forming
protostar.  Much of this chapter will be spent in illustrating that point in
detail.  In addition, {\it chemistry in the first phases of star
formation may affect the chemical composition of the objects that will
eventually form the planetary system}: planets, comets, and
asteroids. In this sense, the study of chemistry during the first
phases of star formation is far-reaching.  Indeed, one of the major
and more fascinating questions linked to the process of the formation
of a star and its planetary system, especially if similar to our Solar
System, is: {\it what is the chemical budget acquired during the
protostellar phase and inherited by the forming planets?}  Answering this
question might even shed some light on our own origins.  The above
ultimate question implies answering several linked questions: what
molecules are formed during the protostellar phase? Do molecules exist
that are formed prevalently during the protostellar phase, and that are
therefore a hallmark of this period?  What is their fate?  Do they
condense onto the grain mantles during the proto-planetary phase?  Are
they incorporated into  planetesimals, which eventually form 
comets, meteorites and planets? Are these pristine molecules released
into nascent  planetary atmospheres during  early, intense cometary
bombardment?  And what is the ultimate molecular complexity reached
during star formation?  In this contribution we will
focus on the first questions, those related to the first phase of 
star formation.

The story starts with the first step towards collapse: the Prestellar
Core phase (see the chapters by {\it Di Francesco et al.} and {\it
  Ward-Thompson et al.}).  These objects are cold ($\leq 10$ K) and
dense ($\geq 10^5$ cm$^{-3}$) condensations inside molecular
clouds. What makes them particularly important is that they are
believed to be on the verge of collapse
\citep{1998ApJ...504..900T,2002ApJ...565..331C,2004ApJ...614..252Y,2005A&A...439.1023C},
and, in this sense, they are considered to be representative of the
initial conditions of star formation.  During this phase, matter
slowly accumulates towards the center under the gravitational force,
which counteracts the thermal and/or magnetic pressure. As the density
increases, gaseous molecules start to freeze-out onto the cold dust
grains, forming H$_2$O-dominated ice mantles ``dirtied'' with several
other molecules. The process is so efficient that in the innermost and
densest regions of the condensation, heavy-element-bearing molecules
are thought to be virtually all frozen out. The low temperatures and
the disappearance of most molecules, and particularly of CO, from the
gas phase trigger a peculiar chemistry: {\it an extreme molecular
  deuteration}.

\begin{table}[tb]
  \caption{Summary of the properties of a sample of Prestellar Cores.
    $n_c(\rm H_2)$ is the central density; $f_{\rm D}$(CO) is the CO
    depletion factor; the Molecular D/H ratio column contains the
    observed N$_2$D$^+$/N$_2$H$^+$ and/or (D$_2$CO/H$_2$CO)$^{1/2}$
    ratio. }
  \begin{tabular}{l c c c c }
    \hline
    Name & $n_c(\rm H_2)$ & $f_{\rm D}$(CO) & Mol. & Ref. \\
           & (10$^5$ cm$^{-3}$)  & & D/H & $^a$\\
        \hline    
L1521F & 3  & 2.5 & 0.01  & 1,2,3 \\
B68    & 1  & 3.4 & 0.03  & 3,4 \\
L1689B & 1  & 4.5 & 0.09  & 3,6,7,8 \\
L183   & 10 & 12  & 0.22  & 3,9 \\
L1544  & 14 & 14  & 0.23  & 3,6,10 \\
L694-2 & 9  & 11  & 0.26  & 3 \\
L429   & 6  & 16  & 0.28  & 3,6,7 \\
OphD   & 3  & 14  & 0.44  & 3,6,7 \\ 
L1709A & 1  & 5.5 & 0.17  & 6,7\\
L1498  & 1  & 7.5 & 0.04  & 3,5 \\
L1517B & 2  & 9.5 & 0.06  & 3,5 \\
    \hline
  \end{tabular}
  $^a$REFERENCES: 1- \cite{2004A&A...414L..53T};
  2- \cite{2001ApJ...547..814H}; 3- \cite{2005ApJ...619..379C};
  4- \cite{2001ApJ...557..209B}; 5- \cite{2004A&A...416..191T};
  6- \cite{2002A&A...389L...6B}; 7- \cite{2003ApJ...585L..55B};
  8- \cite{2002MNRAS.337L..17R}; 9- \cite{2005A&A...429..181P};
  10- \cite{2002ApJ...569..815T}.
  \label{tab:starless}
\end{table}
Once gravitational contraction finally takes over, a protostar is
born, consisting of a central object, which eventually will become a star,
surrounded by an envelope from which the future star accretes
matter. In the beginning, the envelope is so thick that it obscures
the central object, and the Spectral Energy Distribution (SED) is
totally dominated by the cold outer regions of the envelope, with a
temperature lower than $\sim 30$ K. This phase is represented by 
so-called Class 0 sources \citep{2000prpl.conf...59A}.  Most of the envelope
is cold and depleted of heavy-bearing molecules, frozen onto the grain
mantles, exactly as in Prestellar Cores.  However, the presence of
a central source, powered by  gravitational energy, causes the
heating of the innermost regions of the envelope. In these regions the
dust temperature reaches 100 K, causing the evaporation of the grain
mantles formed during the Prestellar-Core phase. The molecules
trapped in the ices are injected back into the gas phase, giving rise
to a rich and peculiar chemistry. Complex organic molecules are found
in these regions, called {\it Hot Corinos}.

In this chapter, we will show that extreme deuteration and Hot Corinos
are specific signatures of the first phases of low-mass star
formation.  The chapter is organized as follows.
We first briefly describe the physical and chemical structures of
Prestellar Cores and Class 0 source (Section \ref{sec:physical}). We then
discuss in detail the deuteration phenomenon (Section
\ref{sec:extreme-deuteration}), Hot Corinos (Section
\ref{sec:hot-corinos}), and the chemical models that
have been developed to explain these observations (Section
\ref{sec:chemical-models}).  The chapter ends with our conclusions.

We conclude this section by listing a few reviews appeared in the
literature related to the subjects treated in this chapter.  Previous
reviews of the chemistry in general are reported in
\cite{2005dmu..conf..205H}, and during the star formation period in
\cite{1998ARA&A..36..317V}, \cite{2000prpl.conf...29L} and
\cite{2005astro.ph..4298C}. Reviews of the observations of multiply
deuterated molecules as well as the related chemical models can be
found in \cite{2003SSRv..106...61R}, \cite{2004dimg.conf..473C} and
\cite{2005A&G....46b..29M}, and on the Hot Corinos in
\cite{2004ASPC..323..195C}.

%%%%%%%%%%%%%%%%%%%%%%%%%%%%%%%%%%%%%%%%%%%%%%%%%%%%%%%%%%%%%%%%%%%%%%%%%
\section{THE PHYSICAL AND CHEMICAL STRUCTURE IN THE FIRST PHASES OF
  THE COLLAPSE}\label{sec:physical}
Since PPIV, our understanding of the physical (temperature and density
profiles) and chemical (molecular and atomic abundance profiles)
structure of matter in the first phases of  low-mass star
formation has improved considerably. This is due both to  the increased
sensitivity of ground-based instruments and to the development of
sophisticated models to interpret the data. In this section, we review
the observations and their interpretations, which  have led to the
reconstruction of the gas density, temperature and molecular
composition in both Prestellar Cores and Class 0 sources.

\subsection{Prestellar Cores}\label{sec:pre-stellar-cores}

\begin{figure}[tb]
 \includegraphics[width=8cm]{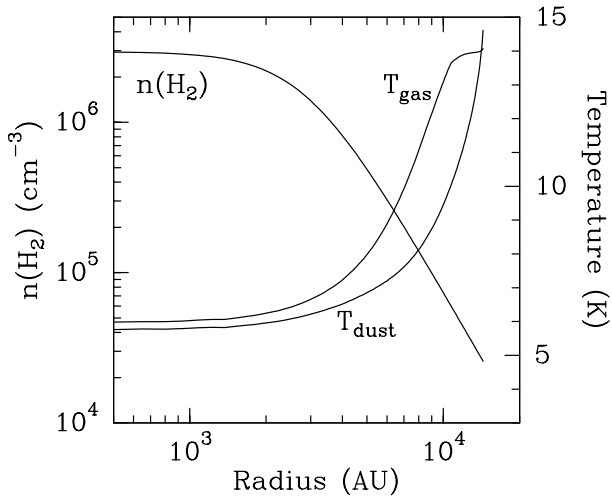}
  \caption{The density and temperature profiles derived for the
  Prestellar Core L1544 (adapted from {\it Crapsi et al.} in prep.).}
  \label{fig:l1544-structure}
\end{figure}

Several authors have studied the physical and chemical structure of
starless cores; detailed reviews are presented in the chapters by {\it Di Francesco et
al.} and {\it Ward-Thompson et al.}. Here we briefly
recall the basic properties characterizing Prestellar Cores, and
discuss some aspects complementary to those presented by the mentioned
authors.  First, the density profiles of the studied Prestellar Cores
are reasonably well represented by Bonnor-Ebert spheres. In practice,
the density profile can be approximated by a power law $r^{-2}$ in the
outer regions and a plateau in the center regions of the Prestellar
Core. In the following we will refer to the density in the plateau as
the ``central'' density.  Typically, the radius of the plateau is
3000-6000 AU \citep{2005ApJ...619..379C}. Figure
\ref{fig:l1544-structure} shows the example of L1544, one of the best
studied Prestellar Cores
\citep{1998ApJ...504..900T,2002ApJ...565..331C,2002ApJ...565..344C},
but other Prestellar Cores present basically the same structure
\citep{2001Natur.409..159A,2005ApJ...619..379C}. In the same figure,
the predicted temperature profile is also shown.  Note the temperature
drop in the inner 3000 AU region, confirmed by observations of NH$_3$
({\it Craspi et al.} in prep). A similar drop in the dust temperature has
been previously observed in other Prestellar Cores
\citep{2001ApJ...557..193E,2003A&A...406L..59P,2004A&A...417..605P,2005A&A...429..181P},
as expected from radiative transfer modeling of centrally concentrated
and externally heated dense cloud cores
\citep{2001A&A...376..650Z}. These large central densities and low
temperatures are accompanied by the depletion of CO molecules in the
central regions, as they condense out onto the grain mantles
\citep{2001ApJ...557..193E,1999ApJ...523L.165C,2002A&A...394..275G}.
Indeed, if one had to give a short list of properties for the
Prestellar Cores they would be large central densities: $\geq 10^5$
cm$^{-3}$; very low central temperatures: $\leq 10$ K; depletion of
molecular species, including CO; and enhanced molecular deuteration.
Table \ref{tab:starless} summarizes these properties in a sample of
Prestellar Cores.
\begin{figure}[bt]
 \includegraphics[width=10cm]{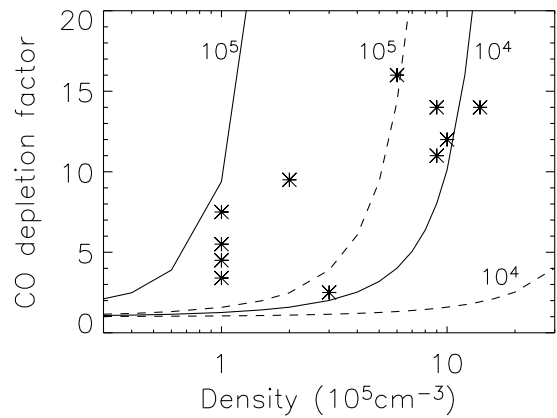}
  \caption{CO depletion factor versus the H$_2$ density. The asterisks
  represent the observations in the Prestellar Cores quoted in Table
  \ref{tab:starless}. The solid lines show the theoretical curves for
  an age of $1\times10^{4}$ and $1\times10^{5}$ yr, respectively,
  assuming an average grain radius of 0.1 $\mu$m. The dashed lines are
  computed taking an average grain radius of 0.5 $\mu$m.}
  \label{fig:psc_plot}
\end{figure}
In Figure \ref{fig:psc_plot} we show the observed CO depletion factors
versus the central densities, where the depletion factor is computed
with respect to the standard CO abundance in molecular clouds,
$9.5\times10^{-5}$ with respect to H$_2$
\citep{1982ApJ...262..590F}. As can easily be seen, there is a clear
correlation between the CO depletion factors and the densities. This
observationally confirms the basic idea that CO molecules disappear
from the gas phase because they freeze-out onto the grains, and that
the condensation rate is proportional to the density (see
below). Figure \ref{fig:psc_plot} also reports theoretical curves,
obtained assuming that the CO molecules condense onto the grains at a
rate $k_{\rm freeze}$ and are released in the gas phase because of the
evaporation caused by cosmic rays at a rate $k_{\rm
cr}=9.8\times10^{-15}$ s$^{-1}$ \citep{1993MNRAS.261...83H}. The
freezing rate can be written as follows:
%
%\begin{equation}
%k_{\rm freeze}=S \left<\pi a_{\rm gr}^{2} n_\mathrm{g}\right> v_\mathrm{CO},
%\label{eq:1}
%\end{equation}
\begin{equation}
k_{\rm freeze}=S \left< \pi a_{\rm gr}^{2} v_\mathrm{CO} \right> n_\mathrm{g} ,
\label{eq:1}
\end{equation}

where we adopted a sticking coefficient $S=1$
\citep{1983ApJ...265..223B}, and a mean grain radius of
$a_\mathrm{gr}$.  The grain number density $n_{\rm g}$ is given by the
(mass) dust-to-gas ratio (0.01) multiplied by the gas density, and
divided by the grain mass (computed assuming a grain density of 2.5 gm
cm$^{-3}$).  The figure shows curves assuming an average grain radius
of 0.1 (solid lines) and 0.5 $\mu$m (dashed lines) respectively. In
the first case, which is the typical value assumed in chemical models
for the ISM, the denser Prestellar Cores lie around the curve with an
age of $1\times10^{4}$ yr, which is definitively too short an age for
these objects, based on several arguments. However, if the average
grain radius is larger and equal to 0.5 $\mu$m, for example, the
observed points lie around the curve at $1\times10^{5}$ yr, which is a
more realistic estimate of the age of these objects. Larger grains
would shift the curve even more to larger ages. Note that we used the
same value for the cosmic ray evaporation rate, although increasing
the grain sizes decreases it because of the larger volume to heat
\citep{2004ApJ...603..159B,2004A&A...415..203S}, enhancing the
effect. Since the surface $\pi a_{\rm gr}^{2}$ is dominated by the
small grains, Figure \ref{fig:psc_plot} suggests that the small grains
are efficiently removed by coagulation with large grains in the
innermost regions of Prestellar Cores
\citep{1993A&A...280..617O,2003A&A...399L..43B,2005A&A...436..933F}. Other
explanations are also possible. For example, the cosmic ray
evaporation rate can be enhanced if the ice mantles contain a
substantial (~1\%) fraction of radicals due to UV photolysis
\citep{1982A&A...109L..12D,1986A&A...158..119D}.  However, in order to
explain Figure \ref{fig:psc_plot} the mantle evaporation rate should
be increased by more than a factor 30 with respect to the
\cite{1993MNRAS.261...83H} rate.  Therefore, our preliminary
conclusion is that, most likely, the average sizes of the grains in
Prestellar Cores are larger than in the ISM, and that these objects
are older than about $1\times10^{5}$ yr.  It is also interesting
noticing that less dense Prestellar Cores point to greater ages and/or
smaller average grain sizes. It is not clear at this stage whether
this means that those cores will not evolve into denser ones, or
whether Prestellar Cores spend more time in this first ``low-density''
phase than in the ``high-density'' phase.

%CC: WE SHOULD MENTION THE N2 PROBLEM HERE!!!!

\subsection{Class 0 sources}\label{sec:class-0-sources}

The physical and chemical structure of the envelopes of Class 0
sources is affected by the presence of a central object, i.e. a
heating source, which implies gradients in the density and
temperature. This, in turn, implies a gradient in the chemical
composition of the gas across the envelope.  Two basic methods have
been used in the literature to derive the chemical and physical
structure of Class 0 envelopes: i) modeling of multi-frequency
observations of several molecules, and ii) modeling of the continuum
SED and map, coupled with molecular multi-frequency observations.  The
first method makes use of the intrinsic property that each line from
any molecule probes the specific region where the line is excited,
namely a region with a specific density and temperature. Using several
lines from the same molecule can, therefore, be used to reconstruct
the {\it gas} density and temperature profiles of the envelope. The
observations of several molecules are then used to derive the
abundance profile of each molecule. Theoretical examples of this
method, applied to the problem of collapsing envelopes, are discussed
in \cite{2000A&A...355.1129C}, \cite{2002Ap&SS.281..139M},
\cite{2003A&A...410..587C}, and \cite{2005A&A...441..171P}.  In these
models the density is assumed to follow the evolution of a Singular
Isothermal Sphere contracting under the hypothesis of isothermal
collapse - the so-called ``inside-out'' framework, developed by Shu
and collaborators \citep{1977ApJ...214..488S}.  In this framework, the
density follows a two-slope power law: in the regions not reached by
the collapse the density has an $r^{-2}$ dependence, whereas in the
collapsing regions the density follows an $r^{-3/2}$ law.
\begin{figure}[bth]
 \includegraphics[width=8cm]{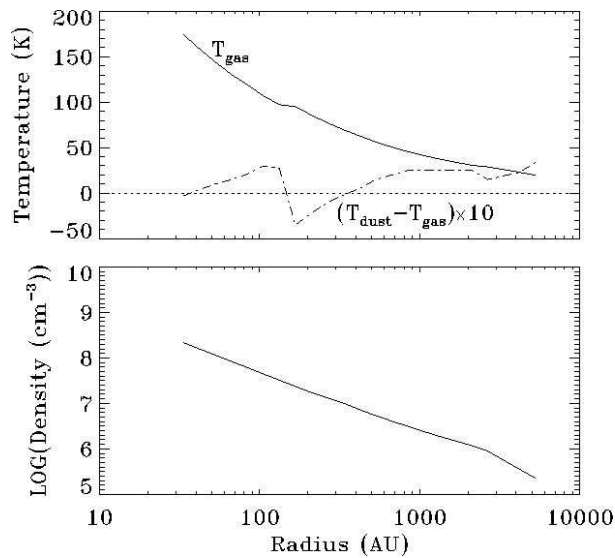}
  \caption{The temperature (upper panel) and density (lower panel)
  profile of the envelope surrounding the Class 0 protostar
  IRAS16293-2422 (figure from \cite{2000A&A...355.1129C}).}
  \label{fig:i16293-structure}
\end{figure}
In addition, in using molecular line observations, one has to also
take into account the abundance profile of the molecule used. The
simplest profile is that of a single step function, which assumes that
the relevant molecule has an almost constant abundance in the cold
outer envelope and a jump in the inner region when the grain mantles
sublimate. Other profiles have been used though, in which the
abundance also has a drop in the outer envelope, to take into account
the molecular depletion as in Prestellar Cores
\citep{2001A&A...372..998C,2004A&A...418..185S,2004A&A...416..603J}.

The second method uses the dust continuum SED simultaneously with maps
to trace the H$_2$ distribution, and the observations are best-fitted
with assumed single power law density distributions
\citep{2002ApJ...575..337S,2002A&A...389..908J,2005A&A...434..257W,2005ApJ...627..293Y}. Then, in order to derive the molecular
abundance profiles, single-jump and/or drop models are used to
interpret multi-frequency observations, as in
\cite{2002A&A...390.1001S}, \cite{2004A&A...418..185S},
\cite{2004A&A...416..603J}, and \cite{2005ApJ...626..919E}. Both
methods give approximatively the same structure (temperature and
density distribution) when applied to the same object, as in the case
of IRAS16293-2422 \citep{2000A&A...355.1129C,2002A&A...390.1001S}.
Figure \ref{fig:i16293-structure} shows the derived structure of the
envelope of this source, obtained by applying the first method to
several molecules \citep{2000A&A...355.1129C,2000A&A...357L...9C}.

Both classes of models predict the existence of regions with dust
temperatures larger than 100 K, the sublimation temperature of the
grain mantles \citep{1996ApJ...471..400C}.  In these regions,
therefore, the components of the grain mantles are injected into the
gas phase, and the abundances of the corresponding molecules should
increase with respect to the abundance in the cold (T$_{dust} \leq
100$ K) envelope.  Water is the major component of the ice mantles,
but unfortunately water lines are not observable from ground-based
telescopes. The spectrometer LWS on board ISO provided observations of
water lines towards a few Class 0 sources \citep{1999A&A...342L..21C},
but with relatively poor spatial and spectral resolution, so that the
interpretation of the data is not unique.  Two interpretations have
been advanced. The first one assumes that the water lines originate in
the shocks at the interface between the outflows and the envelopes of
these sources \citep{2001ApJ...555...40G,2002ApJ...574..246N}. The
second one assumes that the water lines originate mostly in the
envelope. Based only on the ISO data, it is impossible to discriminate
observationally between the two hypotheses. However, if the water line
spectra are interpreted as being emitted in the envelope, allowing a
jump in the water abundance in the T$_{dust}\geq 100$ K region, the
agreement among the model predictions and the observed line fluxes is
rather good \citep{2000A&A...355.1129C, 2002A&A...395..573M}. The
observed data are reproduced if the water abundance is $\sim 3\times
10^{-7}$ in the outer envelope, and jumps by about a factor 10 in the
inner region. In support of this interpretation, the physical
structure derived by this analysis is substantially confirmed by
observations of other molecules obtained with ground-based telescopes
(see below).

Formaldehyde is also an important grain mantle component, and this
molecule has the advantage that it has several transitions in the
millimeter to sub-millimeter wavelength range  observable
with ground-based telescopes
\citep{1993ApJS...89..123M,2003A&A...410..587C}. These observations
have a much better spatial and spectral resolution than ISO-LWS and
provide stronger constraints for the models. A survey of the H$_2$CO
line emission towards a bit fewer than a dozen Class 0 sources has
shown that all the targeted sources, except VLA16293, have a region
where the formaldehyde abundance jumps by more than one order of
magnitude \citep{2000A&A...357L...9C,2004A&A...416..577M}. Several of
these sources also show jumps in the methanol abundance
\citep{2005astro.ph..7172M}. The {\it predicted} sizes of the jump
regions range from 10 to 150 AU, and are, therefore, comparable to the
sizes of the Solar System. The densities are predicted to be larger
than 10$^8$ cm$^{-3}$. However, one should be aware that the above
observations have been obtained with {\it single-dish} telescopes,
which, at best, have spatial resolutions corresponding to about 1000
AU in radius. Therefore, both the estimates of the abundances and the
warm region sizes suffer from a relatively large uncertainty. Besides,
the details of the model adopted in the analysis of the data give rise to
additional systematic uncertainties so that the very existence of the
H$_2$CO abundance jumps is contested by some authors
\citep{2004A&A...418..185S,2005A&A...437..501J,2005ApJ...632..973J}. In
addition, the situation is further confused by the presence at small
scales of cavities and outflows, which some authors also consider a
major component in the line emission attributed to the abundance jumps
of the above analysis \citep{2005ApJ...632..371C,2005ApJ...631L..77J}.

Finally, several simple molecules ( HCO$^+$, N$_2$H$^+$, CS, SO,
SO$_2$, HCN, HNC, HC$_3$ N and CN) have been observed in a sample of
Class 0 sources \citep{2004A&A...416..603J}. Given the observed
transitions, this study was able to trace the abundance of the
observed species in the outer envelope and probe some formation
routes, such as for N$_2$H$^+$ and HCO$^+$. One remarkable result is
the discovery of large regions in the outer envelopes where the
molecular abundances are depleted, because of condensation onto 
grain mantles \citep{2004A&A...416..577M,2005A&A...435..177J}.
However, even more remarkable is the discovery of
extreme deuterium fractionation in the Class 0 sources, which will be
discussed in Section \ref{sec:extreme-deuteration}, and which is indeed
linked to the molecular depletion.\\

In summary, from a chemical point of view, the similarity to
Prestellar Cores is the most prominent characteristic of the outer
envelopes of Class 0 sources. With respect to molecular depletion and
extreme deuteration, they are indeed virtually indistinguishable,
suggesting that Prestellar Cores are the likely precursors of Class 0
sources \citep{2004A&A...416..577M,2005A&A...437..501J}.

%%%%%%%%%%%%%%%%%%%%%%%%%%%%%%%%%%%%%%%%%%%%%%%%%%%%%%%%%%%%%%%%%%%%%%%%%
\section{EXTREME DEUTERATION}\label{sec:extreme-deuteration}
%\bigskip

The previous section has anticipated a major characteristic shared by
Prestellar Cores and the envelopes of Class 0 sources: an extreme
molecular deuteration. Although the deuterium abundance is
$1.5\times10^{-5}$ relative to hydrogen \citep{2003SSRv..106...49L},
singly deuterated molecules have been observed in both types of
objects with abundances relative to their hydrogenated counterparts
between 10\% and 50\%
\citep{1977ApJ...217L.165G,1990ApJ...365..269L,1995ApJ...447..760V,2000A&A...356.1039T,2001ApJ...554..933S,2002P&SS...50.1173R,2004A&A...420..957C,2005ApJ...619..379C}. Even
more extreme, doubly and triply deuterated molecules have been
observed (Table \ref{tab:psc:deut}) with D/H ratios reaching 30\% for
D$_2$CO \citep{2002P&SS...50.1205L} and 3\% for CD$_3$OH
\citep{2004A&A...416..159P}. This enhances the molecular D/H ratio by
up to 13 orders of magnitude with respect to the elemental D/H
ratio. In this section, we review these spectacular observations,
which have led to the development of a new class of models for
molecular deuteration, described in Section \ref{sec:chemical-models}.  It
is worth emphasizing that extreme fractionations have only been
observed in {\it low mass} Prestellar Cores (Section
\ref{sec:deut:pre-stellar-cores}) and protostars (Section
\ref{sec:deut:class-0-sources}) so far. Massive protostars do not show
the same deuterium enrichment, probably because the chemistry occurs
in warmer environments.

\subsection{Prestellar Cores}\label{sec:deut:pre-stellar-cores}

In the last few years, it has become increasingly clear that molecular
deuteration reaches extreme values in Prestellar Cores, where multiply
deuterated molecules have been detected with ratios larger than
1\%. The list of multiply deuterated molecules detected in Prestellar
Cores (and Class 0 sources) is reported in Table \ref{tab:psc:deut},
as well as the typical measured D/H ratio for each species.
\begin{table}[tb]
  \caption{Summary of the multiply deuterated molecules observed in
    Prestellar Cores and Class 0 sources. Typical deuterium
    fractionation ratios are given with respect to the fully hydrogenated
    counterparts, except for D$_2$S and HD$_2^+$ where the ratio is
    with respect to the singly deuterated isotopologue. Note that
    variations among different sources amount to about a factor 3 or
    so.}
    \begin{tabular}{l c l}
    \hline
    Species & Deuterium     & Ref. \\
            & Fractionation &  $^a$\\
        \hline    
NHD$_2$   & 0.03  & 1,2,3\\
 ND$_3$   & 0.006 & 3,4,5\\
D$_2$CO   & 0.15  & 6,7,8\\
D$_2$S    & 0.12  & 9\\
HD$_2^+$  & 1     & 10\\
CHD$_2$OH & 0.06  & 11,8\\
CD$_3$OH  & 0.03  & 12\\
D$_2$CS   & 0.10  & 13\\
    \hline
  \end{tabular}

   $^a$REFERENCES:1- \cite{2000A&A...354L..63R}; 2-
    \cite{2001ApJ...552L.163L}; 3- \cite{2005A&A...438..585R} 4-
    \cite{2002ApJ...571L..55L}; 5- \cite{2002A&A...388L..53V}; 6-
    \cite{1998A&A...338L..43C}; 7- \cite{2003ApJ...585L..55B}; 8-
    \cite{Parise06}; 9- \cite{2003ApJ...593L..97V}; 10-
    \cite{2004ApJ...606L.127V}; 11- \cite{2002A&A...393L..49P}; 12-
    \cite{2004A&A...416..159P}; 13- \cite{2005ApJ...620..308M}.
  \label{tab:psc:deut}
\end{table}
To the best of our knowledge, two surveys of multiply deuterated molecules --
 D$_2$CO
\citep{2003ApJ...585L..55B} and ND$_3$ \citep{2005A&A...438..585R} --
have been obtained in (small) samples of Prestellar Cores.
\cite{2003ApJ...585L..55B} found that the D$_2$CO/H$_2$CO ratios
correlate with the CO depletion factors, proving observationally the
key role of CO depletion in the deuterium enhanced fractionation. This
trend is confirmed now by the larger sample of Table
\ref{tab:psc:deut}, as shown in 
\begin{figure}[bt]
 \includegraphics[width=10cm]{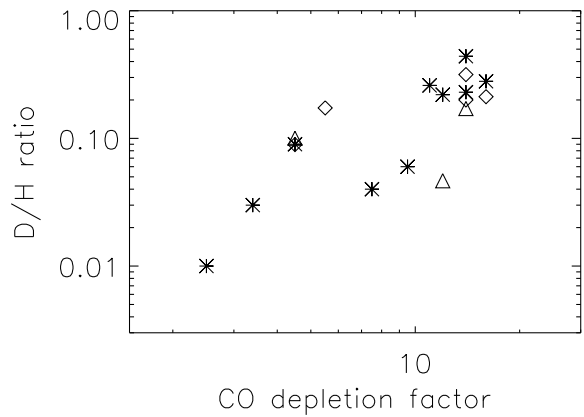}
  \caption{The measured N$_2$D$^+$/N$_2$H$^+$ (stars; from
    \cite{2005ApJ...619..379C}), (D$_2$CO/H$_2$CO)$^{1/2}$ (diamonds;
    \cite{2003ApJ...585L..55B}) and (ND$_3$/NH$_3$)$^{1/3}$
    (triangles; \cite{2005A&A...438..585R}) in the sample of
    Prestellar Cores of Table \ref{tab:psc:deut}, as function of the
    CO depletion factor.}
% the point with different ND3 and N2H+ is L183, but it could be due
%    to a different pointing
\label{fig:psc_dh}
\end{figure}
Figure \ref{fig:psc_dh}: {\it the larger the CO depletion the larger
the molecular D/H ratio} (see also \cite{2005ApJ...619..379C}). This
is a key point whose importance will be further discussed in Section
\ref{sec:chemical-models}. Another aspect shown in the figure is the
similarity of the D/H ratio of the three observed molecules:
N$_2$H$^+$, H$_2$CO and NH$_3$. Note that the three molecules refer to
the D/H ratios of singly, doubly and triply deuterated isotopologues
respectively, and the ratios shown in the figure are scaled to account
for the number of D atoms. While N$_2$H$^+$ is certainly a gas phase
product (formed by the reaction of N$_2$ with H$_3^+$), there is
considerable debate on whether formaldehyde and ammonia are formed in
the gas phase or on grain surfaces
\citep{1983A&A...119..177T,2001A&A...372..998C,2005A&A...438..585R};
Section \ref{sec:chemical-models}). From a purely observational point of
view, Figure \ref{fig:psc_dh} points to a similar origin for the three
molecules, but this is very likely an over-simplification, and
maybe just due to the common origin of the molecular deuteration, the
molecular ion H$_2$D$^+$ (Section \ref{sec:chemical-models}).

Indeed, it has been long believed that the deuterium fractionation is
started by the fast ion-neutral reaction H$_3^+$+HD leading to
H$_2$D$^+$ (Section \ref{sec:molec-deut-gas}). For this reason, H$_2$D$^+$
has long been searched for in regions of massive star formation,
where, however, it has never been detected
\citep{1992A&A...261L..13V,1992A&A...258..472P}. The first detection
of H$_2$D$^+$ was finally obtained by \cite{1999ApJ...521L..67S}
towards the {\it low-mass} Class 0 source NGC1333-IRAS4.  However, the
break-through detection is considered to be that obtained in the
Prestellar Core L1544 \citep{2003A&A...403L..37C}, where the ground
ortho-H$_2$D$^+$ line at 372 GHz is ten times brighter than in
NGC1333-IRAS4. The observed flux implies an abundance of H$_2$D$^+$
similar to that of electrons, and, therefore, to that of H$_3^+$. This
was unexpected and unpredicted by the chemical models at the time of
the discovery. It is now accepted that the H$_2$D$^+$/H$_3^+$ ratio
can reach unity in cold gas depleted by heavy-element-bearing
molecules, as occurs at the centers of Prestellar Cores (Section
\ref{sec:pre-stellar-cores}). Actually, in the extreme conditions of
the centers of Prestellar Cores, even the doubly deuterated form of
H$_3^+$ has been detected \citep{2004ApJ...606L.127V}. The map of
H$_2$D$^+$ emission in L1544 shows that it is very abundant in a zone
with diameter $\sim6000$ AU \citep{2006vastel}. This zone coincides
with the central density plateau, where CO is highly depleted
\citep{1999ApJ...523L.165C}. This further proves the key role played
by the freezing of heavy-element-bearing molecules in the enhancement
of molecular deuteration. Finally, one has to note that L1544 is not a
peculiarity. An on-going project shows that Prestellar Cores are in
general strong emitters of the ground ortho-H$_2$D$^+$ line ($T_{\rm
MB}$ $>$ 0.5 K; {\it Caselli et al.}, in prep.).

Thanks to the above observations,   the very high
diagnostic power of the ground-state ortho-H$_2$D$^+$ line is now realized. This is
true not only for chemistry, but also for the study of the
evolutionary and dynamical status of Prestellar Cores.  For
example, the ortho-to-para ratio of H$_2$D$^+$ is predicted to
critically depend on the density, temperature and grain sizes
\citep{2005A&A...436..933F,2006astro.ph..1429F}. Comparing these model
predictions with observations, \cite{2006vastel} found that the
average grain sizes in L1544 are larger than in the ISM (in agreement
with the discussion in Section \ref{sec:pre-stellar-cores}).  Finally,
since very likely most of the heavy-element-bearing molecules are
frozen out onto the grain mantles in the centers of the Prestellar
Cores, H$_2$D$^+$ is the best, if not the only way to probe those
regions. Of particular interest is knowing the kinematical structure
and whether and when the collapse  starts at the
center. \cite{2005A&A...439..195V} studied the lines profile of the
fundamental ortho-H$_2$D$^+$ line towards L1544 with this scope. They
found that  the infall likely started at the center of L1544. We
predict that similar studies with submillimeter interferometers will
be extremely useful to unveil the first instants of collapse.

%\bigskip
%\noindent
\subsection{Class 0 sources}\label{sec:deut:class-0-sources}

As for Prestellar Cores, recent observations have revealed a zoo of
molecules with enhanced deuterium fractionation. Actually, the
discovery of abundant D$_2$CO in a Class 0 source started the hunt for
multiply deuterated molecules.  Although doubly deuterated
formaldehyde had been detected fifteen years ago in Orion by
\cite{1990ApJ...362L..29T}, the measured abundance ($\sim 0.2$\%
H$_2$CO) did not seem to draw particular attention (probably because
Orion has always been considered a peculiar region). The discovery
that D$_2$CO has an abundance about 5\% that of H$_2$CO in a low mass
protostar, IRAS16293-2422 \citep{1998A&A...338L..43C}, on the
contrary, started a flurry of activity in this field, and initiated
the hunt for multiply deuterated molecules.  In a few years, the
deuterated species quoted in Table \ref{tab:psc:deut} were detected.

\begin{figure}[tb]
 \includegraphics[width=9cm]{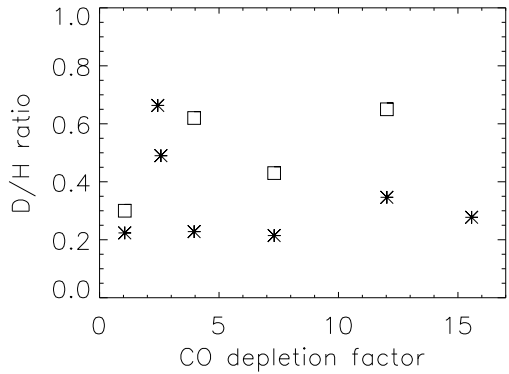}
  \caption{The measured (D$_2$CO/H$_2$CO)$^{1/2}$ (stars) and
    CH$_2$DOH/CH$_3$OH (squares) in a sample of Class 0 sources
    \citep{Parise06}\, as function of the CO depletion factor.}
\label{fig:cl0-deuteration}
\end{figure}
The best studied Class 0 source is IRAS16293-2422, where
multi-frequency observations of singly and doubly deuterated
formaldehyde constrain the abundance ratios of these molecules quite
well \citep{2000A&A...359.1169L} to be HDCO/H$_2$CO = 0.15 and
D$_2$CO/H$_2$CO = 0.05. Besides, D$_2$CO is very abundant across the
entire envelope, which extends up to a radius of more than 5000 AU
\citep{2001A&A...372..998C}. Remarkably, the D$_2$CO/H$_2$CO ratio
reaches a peak of 0.16 at about 2000 AU from the center. Note that the
dust at this distance is warm enough to prevent the condensation of
CO, and the gas is too warm, $\sim35$ K, to favor the deuterium
fractionation in the gas phase
\citep{2000A&A...355.1129C,2002A&A...390.1001S}.  In a few words, the
two key factors for the enhanced fractionation, cold and CO-depleted
gas, are absent in the largest fraction of the IRAS16293-2422
envelope, where D$_2$CO is abundant. Therefore, the observed enhanced
deuterium fractionation is unlikely to be a present-day product
\citep{2001A&A...372..998C}. In support of this view, the formaldehyde
abundance profile across the IRAS16293-2422 envelope increases by
about a factor of ten where the dust temperature exceeds 50 K,
suggesting that gaseous formaldehyde is due to the partial, but
continuous sublimation of the grain mantles across the entire
envelope. Layered like onions, the components of the mantles sublimate
at different distances, corresponding to different dust
temperatures. In the regions where the dust temperature is larger than
about 25 K, CO-rich mantles sublimate, injecting into the gas phase
the CO and the trace molecules trapped in the ice, like
formaldehyde. When the dust temperature reaches 50 K the H$_2$CO-rich
ices sublimate and so on \citep{2001A&A...372..998C}. {\it The
observed deuteration is therefore a heritage of the pre-collapse
phase} (Section \ref{sec:deut:pre-stellar-cores}).

This can also be seen in Figure \ref{fig:cl0-deuteration}, where the
measured D/H ratios of formaldehyde and methanol are reported as
function of CO depletion in a sample of Class 0 sources
\citep{Parise06}. Contrary to the Prestellar Cores, there is no clear
correlation between the molecular deuteration and the {\it present} CO
depletion. The most plausible explanation is that the deuterated
formaldehyde and methanol have been built up during the previous phase
and that Class 0 sources are young enough for it not being
substantially modified, likely younger than $\sim 10^5$ yr based on
modeling \citep{1997ApJ...482L.203C}.  Figure
\ref{fig:cl0-deuteration} also shows that methanol is systematically
more enriched with deuterium than is formaldehyde. This may reflect a
different epoch of formation of these two molecules on the ices: the
formaldehyde perhaps formed in an earlier stage than methanol, when CO
depletion was less severe. Alternatively, present day gas-phase
reactions may have changed the initial mantle composition already, or,
finally, formaldehyde and methanol follow different routes of
formation (gas phase versus grain surface?)  \citep{Parise06}.
 
Although our comprehension of the mechanisms leading to molecular
deuteration in the ISM has undoubtedly improved in these last few
years, there still remain a number of unresolved questions. First, the
relative role of gas phase versus grain surface chemistry is
uncertain: it is not completely clear yet what molecules and to what
extent are formed in the gas phase versus on the grain surfaces.
Deuterium fractionation promises to be a key aspect in this
issue. Another open question is represented by deuterated
water. Searches for solid HDO have so far been in vain, giving some
stringent upper limits to the water deuteration: $\leq 2$\%
\citep{2003A&A...399.1009D,Pariseetal2003a}, considerably less than
the deuteration observed in the other molecules. One possible
explanation is that observations of solid H$_2$O and HDO may be
``contaminated'' by the contribution of the molecular cloud in front
of the observed protostar (see for example the discussion in
\cite{2002ApJ...570..708B}). A more stringent constraint is set by
observations of gas-phase HDO in the outer envelope and hot corino,
where the icy mantles sublimate. Observations towards IRAS16293-2422
set a stringent constraint to the deuteration of water; in the
sublimated ices it is $\sim3$\%, whereas in the cold envelope it is
$\leq 0.2$\% \citep{2005A&A...431..547P}. Therefore, water deuteration
seems to follow a different route from formaldehyde, methanol, and the
other molecules.
%%CC: D2O?

%%%%%%%%%%%%%%%%%%%%%%%%%%%%%%%%%%%%%%%%%%%%%%%%%%%%%%%%%%%%
%
\begin{figure*}[tbh]
 \includegraphics[width=6cm,angle=90]{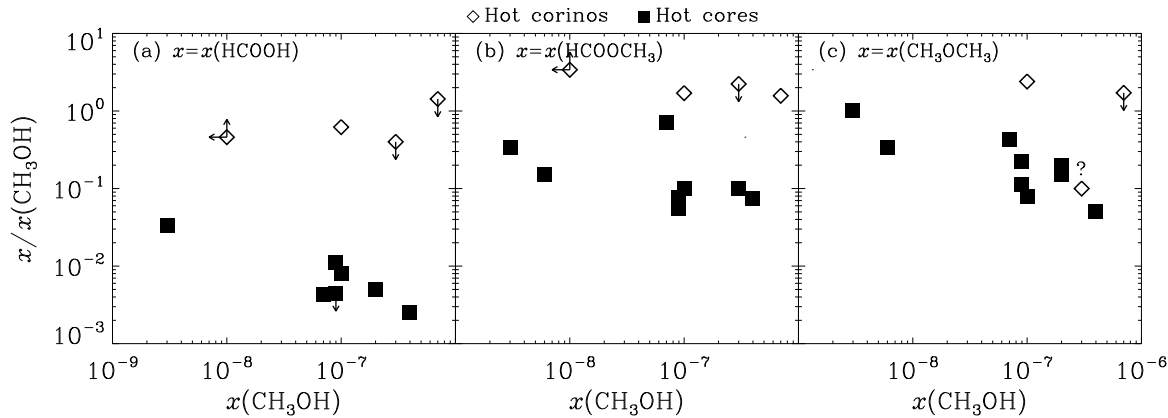}
  \caption{Abundance ratios of complex O-bearing molecules to
    methanol, plotted as a function of the methanol
    abundance. Diamonds and filled squares represent hot corinos and hot
    cores respectively (from \cite{bottinelli2005submitted}).}
  \label{fig:hotcorinos}
\end{figure*}
\section{HOT CORINOS}\label{sec:hot-corinos}

While the outer envelopes of Class 0 sources are so cold that
heavy-element-bearing molecules are at least partially frozen onto the
grain mantles, the innermost regions are characterized by warm dust,
so warm that the mantles sublimate totally (Section
\ref{sec:class-0-sources}). These regions share several aspects of the
hot cores, discovered in the 80s around massive protostars
\citep{2000prpl.conf..299K}. Being ``smaller'' in size and mass, they
have been called {\it hot corinos}
\citep{2004ASPC..323..195C,2004ApJ...617L..69B}. Based on the analysis
discussed in Section \ref{sec:class-0-sources}, the radii of the hot
corinos are predicted to be around 50 AU (IRAS16293-2422 represents
here an exception with a radius of 150 AU), comparable to the sizes of
the Solar System.  The key question, and the most relevant for the
chemical structure of Class 0 sources, is which molecules are found in
the hot corinos, and, in particular, whether complex organic molecules
are formed, in analogy with massive hot cores. More generally, are the
hot corinos a scaled version of the hot cores, or does a peculiar
chemistry take place? And, how reliable are the hot corino sizes
derived by the analysis of single-dish data?

Once again, the benchmark for these studies has been IRAS16293-2422.
In the hot corino of this source, complex organic molecules were
detected for the first time \citep{2003ApJ...593L..51C}, and this hot
corino was the first to be imaged
\citep{2004ApJ...616L..27K,2004ApJ...617L..69B}.  In practice, all the
complex molecules typical of massive hot cores that have been searched
for were detected, with abundances similar to if not larger than those
measured in hot cores \citep{2003ApJ...593L..51C}.  These species
include O and N-bearing molecules such as formic acid, HCOOH,
acetaldehyde, CH$_3$CHO, methyl formate, HCOOCH$_3$, dimethyl ether,
CH$_3$OCH$_3$, acetic acid, CH$_3$COOH, methyl cyanide, CH$_3$CN,
ethyl cyanide, C$_2$H$_5$CN and propyne, CH$_3$CCH.  Two years after
the first detection in IRAS16293-2422, complex organic molecules were
also detected towards the other Class 0 sources NGC1333 IRAS4A and B,
and NGC1333 IRAS2
\citep{2004ApJ...615..354B,2005ApJ...632..973J,bottinelli2005submitted}.
The abundances of the O-bearing complex molecules detected in the
four hot corinos are shown in Figure \ref{fig:hotcorinos} as functions
of the methanol abundance \citep{bottinelli2005submitted}. Note that
the abundances have been normalized to the methanol abundance so that
they do not depend on the uncertain sizes of the hot corinos.  The
number of hot corinos where complex organic molecules have been
detected is evidently much too small to try to draw firm conclusions
based on statistical considerations. Nevertheless, it is obvious that
complex organic molecules are rather common in hot corinos. In this
respect, the prototypical Class 0 source, IRAS16293-2422, may actually
be representative for Class 0 sources. Furthermore, the abundances of
HCOOH, HCOOCH$_3$ and CH$_3$OCH$_3$ in hot corinos are comparable to
the methanol abundance and to formaldehyde
\citep{bottinelli2005submitted}, and do not depend appreciably on the
source luminosity, nor on the methanol and formaldehyde abundance. If
the O-bearing complex molecules are formed in the gas phase from
methanol and formaldehyde, as some theories predict (Section
\ref{sec:mantle-evap-compl}), then they ``burn'' the majority of their
``parent'' molecules, and this burned fraction is independent of the
amount of H$_2$CO and/or CH$_3$OH sublimated from the grain
mantles. Alternatively, if the O-bearing complex molecules are
synthesized on the grain surfaces, they are an important ice
component. So far, however, only HCOOH has been claimed to be found in
its solid form \citep{1999A&A...343..966S,2000ApJ...536..347G}.  The
other evident message of Figure \ref{fig:hotcorinos} is that hot
corinos are not a simply scaled version of hot cores. The abundance of
the O-bearing complex molecules in hot corinos is larger than that in
hot cores, by more than a factor of ten: HCOOH is even 100 times more
abundant.

So far, we have focused on results based on single-dish
observations. These observations encompass regions of at least 1000
AU, whereas the predicted sizes of the hot corinos do not exceed 150
AU.  Recently, interferometric observations have resolved the emission
of complex organic molecules in the hot corinos towards IRAS16293-2422
\citep{2004ApJ...616L..27K,2004ApJ...617L..69B,2005ApJ...632..371C},
NGC1333-IRAS2 \citep{2005ApJ...632..973J}, and NGC1333-IRAS4B
({\it Bottinelli et al. }, in prep.).  From these data, a few general
conclusions can be drawn. First, for IRAS16293-2422 and
NGC1333-IRAS4B, the line emission from the complex molecules is
concentrated in two spots, centered on the two objects forming in each
case a proto-binary system
\citep{1989ApJ...337..858W,2003ApJ...592..255L,2001ApJ...546L..49S,2001ApJ...562..770D}. The
full line flux of the observed complex molecules, detected with the
single-dish, is recovered in these two spots, which means that all the
single dish emission originates in those regions. In addition, there
is no sign of outflows in the line emission of complex molecules; the
emission is compact and the line profile is relatively narrow. The
brightest -- and so far only resolved spot -- in IRAS16293-2422 has a
radius of $\sim$ 150 AU, close to the size expected for warm ($T>100$
K) gas \citep{2000A&A...355.1129C}.  As emphasized above, each of
these two objects is indeed a protobinary system. In each case, the
two objects in the protobinary have different characteristics; eg.,
one spot is bright in the continuum but barely detected in the complex
molecule line emission, while this is reversed for the other
spot. This difference does not necessarily imply that the abundances
in the individual spots of these protobinaries are truly
different. The spot with the weaker line emission is also unresolved
and the line likely optically thick \citep{2004ApJ...617L..69B}.
Evidently, high-resolution observations of complex organic molecules
are a valuable tool to study the chemistry of the hot corinos, and how
it depends on physical conditions, rather than the mantle composition
and age, which presumably are the same for the two objects of the
binary system.

%%%%%%%%%%%%%%%%%%%%%%%%%%%%%%%%%%%%%%%%%%%%%%%%%%%%%%%%%%%%%%%%%%%%
\section{CHEMICAL MODELS}\label{sec:chemical-models}

\subsection{Molecular deuteration in the gas phase}\label{sec:molec-deut-gas}

Although the basic gas-phase mechanism of molecular deuteration in the
cold interstellar medium was elucidated almost thirty years ago
\citep{1977ApJ...217L.165G,1978ApJ...222L.145W} and incorporated into
detailed models over a decade ago \citep{1989ApJ...340..906M}, it is
only recently that an understanding of the effect in star-forming
regions has been achieved by astrochemists.  Since the reservoir of
deuterium lies in the molecule HD, deuteration proceeds by removing
deuterium atoms from this molecule onto the many other trace
constituents of the gas.  The principal mode of transfer at low
temperatures has been thought to be the reaction system
\begin{equation}
{\rm H_{3}^{+}  +  HD \rightleftharpoons H_{2} D^{+}  +  H_{2}},
\end{equation}
in which the left-to-right reaction is exothermic by approximately 230
K.  If no other reactions need be considered to determine the
abundance ratio of H$_{2}$D$^{+}$ to H$_{3}^{+}$, the slowness of the
backwards reaction leads to the stunning prediction that the abundance
of the deuterated ion, now referred to as an isotopologue, can exceed
that of H$_{3}^{+}$.  For cold cores at the standard density of
10$^{4}$ cm$^{-3}$, the prediction is in error because H$_{2}$D$^{+}$
reacts quickly with both electrons and heavy species such as CO.  The
reactions with heavy species serve to spread the enhanced deuteration
around so that typical ratios between singly-deuterated isotopologues
(e.g., DCO$^{+}$) and normal species (e.g. HCO$^{+}$) are predicted to
lie in the range 0.01--0.10, in agreement with observation.  Two other
exchange reactions are of lesser importance: these involve the ions
CH$_{3}^{+}$ and C$_{2}$H$_{2}^{+}$.  Their deuterium exchange
reactions with HD leading to the ions CH$_{2}$D$^{+}$ and
C$_{2}$HD$^{+}$ are more exothermic than the H$_{3}^{+}$ case, so that
the backwards reactions are slow up to higher temperatures, and
subsequent reactions involving the deuterated ions can maintain
deuterium fractionation when H$_{2}$D$^{+}$ is no longer enhanced.
Other reactions are also non-negligible, including some involving
deuterium atoms \citep{1989MNRAS.237..661B}.

As the density increases, the picture changes dramatically.  Near the
center of a cold pre-stellar core, for example, the gas density
increases to 10$^{5-6}$ cm$^{-3}$ and the time scale for accretion of
species onto dust particles becomes so short that most heavy species
are strongly depleted onto dust particles. In addition, the fractional
ionization in the gas is thought to become quite low if one takes
account of gas-grain interactions to deplete the chemically rather
inert atomic ion H$^{+}$ \citep{2003A&A...398..621L}.  Since seemingly
all destruction mechanisms of H$_{2}$D$^{+}$ are slow, a large
abundance of H$_{2}$D$^{+}$ does indeed build up, in agreement with
observations (Section \ref{sec:deut:pre-stellar-cores}).  But the story
does not end here: in analogy to the reaction above, further reactions
with HD convert a significant portion of the H$_{2}$D$^{+}$ to the
more deuterated isotopologues HD$_{2}^{+}$ and even D$_{3}^{+}$
\citep{2003ApJ...591L..41R,2004A&A...418.1035W,2005A&A...440..583C}.
Indeed, simple chemical models of the densest portions show the triply
deuterated ion to be the most abundant of the four ions in the series,
and possibly the most abundant ion in the gas
\citep{2003ApJ...591L..41R}.
\begin{figure}[tb]
 \includegraphics[width=9cm]{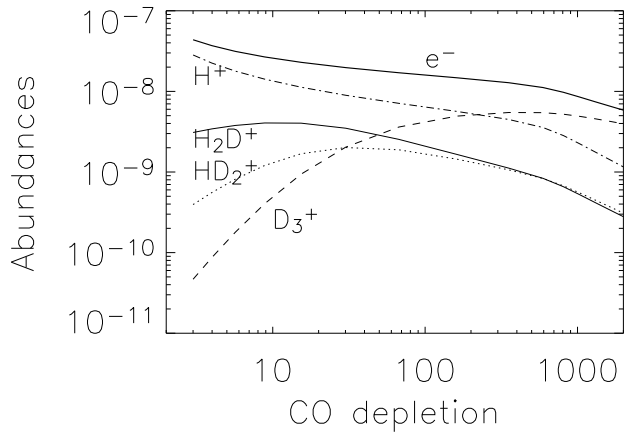}
  \caption{Abundance of the deuterated forms of H$_3^+$, H$^+$ and
  e$^-$, as functions of the CO depletion  factor: thick solid
  line e$^-$, thin solid line H$_2$D$^+$, dashed line D$_3^+$, dotted
  line HD$_2^+$, and dotted-dashed line H$^+$.}
\label{fig:deuteration}
\end{figure}
Figure \ref{fig:deuteration} shows the abundances of the deuterated
forms of H$_3^+$, H$^+$ and e$^-$, as functions of the CO depletion
factor respectively. Since the age and any other parameter is
constant, increasing the depletion factor corresponds to increasing
the density. For ``intermediate'' ($\leq 30$) CO depletion factors,
H$_2$D$^+$ is the most abundant deuterated form of H$_3^+$. However,
for larger depletions D$_3^+$ takes over, and becomes the dominant
positive charge carrier.  A similar effect leads to the ions
CD$_{3}^{+}$ and C$_{2}$D$_{2}^{+}$.

Although a tiny region in the center may be completely devoid of heavy
species \citep{2004A&A...418.1035W}, such molecules do exist, albeit
with depleted abundances, near the center
\citep{2002ApJ...570L.101B,2005A&A...429..181P,2006ApJ...636..916L}.
The remaining heavy species will react with all three of the
deuterating ions, especially D$_{3}^{+}$, leading to highly-deuterated
isotopologues such as D$_{2}$CO and ND$_{3}$.  Critical here is the
high efficiency of D$_{3}^{+}$ since, unlike the partially deuterated
analogs, it can $only$ deuterate and not protonate heavy species.
Contrast the reaction
\begin{equation}
{\rm D_{3}^{+}  +  CO   \longrightarrow  DCO^{+}  +  D_{2}}
\end{equation}
with
\begin{equation}
{\rm  H_{2}D^{+}  +  CO  \longrightarrow DCO^{+} + H_{2}; HCO^{+} +  HD}.
\end{equation}
In the latter reaction, deuteration occurs on only one of three
collisions.  The increase in efficiency with D$_{3}^{+}$ as the
deuterating agent is especially important in the formation of multiply
deuterated neutral isotopologues, which occurs via a cycle of
deuteration followed by dissociative recombination reactions.  For
example, starting with normal ammonia, deuteration leads to
NH$_{3}$D$^{+}$, followed by dissociative recombination to form
NH$_{2}$D, followed by deuteration to form NH$_{2}$D$_{2}^{+}$,
followed by dissociative recombination to form NHD$_{2}$, etc
\citep{2005A&A...438..585R}.

Gas-phase chemical models of Prestellar Cores including deuteration
range from simple homogeneous treatments relevant to the center of the
core \citep{2003ApJ...591L..41R} to shell models that can be static
\citep{2004A&A...424..905R} or in a state of collapse
\citep{2003ApJ...593..906A,2005ApJ...620..330A}.  Accretion onto
grains must be included in all approaches.  The collapse can be
treated via analytical approaches \citep{2003ApJ...593..906A} or
hydrodynamic simulations \citep{2005ApJ...620..330A}.  The chemistry
occurring on grain surfaces can also be included
\citep{2003ApJ...593..906A,2005ApJ...620..330A}. A relatively simple
static approach by \cite{2004A&A...424..905R} illustrates the power
of the chemical models. In terms of column densities,
\cite{2004A&A...424..905R} obtain a value for D$_{2}$CO higher by
only a factor of three than that observed by
\cite{2003ApJ...585L..55B}, and a CO depletion of a factor of 20
compared with the standard value, in good agreement with the observed
depletion of 14 \citep{2002A&A...389L...6B} (see also Section
\ref{sec:extreme-deuteration} and \ref{sec:physical}).  The overall
computed D$_{2}$CO/H$_{2}$CO ratio is 0.3, also in good agreement with
observation.  In general, models with collapse are not in as good
agreement with observation because it is not facile to vary the
details of the collapse to obtain optimal agreement.

% INCLUDE  roueff's models!

\subsection{Mantle formation and surface chemistry}\label{sec:mantle-form-surf}

As the gas-phase chemistry occurs and mantles accrete in cold sources,
a surface chemistry is also active.  A proper mathematical treatment
of diffusive surface reactions on interstellar grains is not simple.
The first such treatments used rate equations analogous to those used
for gas-phase processes \citep{1977Ap&SS..52..443P}, and the method is
still the only practical approach to the construction of large
gas-grain models \citep{2000MNRAS.319..837R}.  The method is only a
rough approximation to reality for a number of reasons, and a variety
of more accurate but more complex methods have been advocated.  If we
imagine grains to be rather smooth surfaces on which binding
parameters for adsorbates do not differ strongly from one site to
another, there is the problem that only small (fractional) numbers of
reactive species such as atomic hydrogen are present on a given grain
on average.  With such small numbers, it is more reasonable to
consider discrete rather than average quantities. Both the
discreteness of the problem and the large fluctuations possible from
grain to grain then argue for a stochastic approach.  Following the
pioneering work of \cite{1982A&A...114..245T}, a truly stochastic
method was utilized by \cite{2001ApJ...562L..99C} based on Monte Carlo
methods, while a master equation approach was advocated by
  \cite{2001ApJ...553..595B} and by \cite{2001A&A...375.1111G}.  The
latter approach has recently been incorporated into moderately sized
gas-grain networks by \cite{2002PhRvE..66e6103B}, \cite{2004PhRvL..93q0601L}
and \cite{2004A&A...423..241S}.

To add to the complexity of the problem, interstellar surfaces are
likely to be irregular, porous, and even amorphous, so that binding
sites for adsorbates are likely to be diverse, including some that
contain much deeper wells than others.  A new extension of the Monte
Carlo method for irregular grains with silicate, carbonaceous and icy
surfaces has been introduced by \cite{2005A&A...434..599C} and
\cite{2005MNRAS.361..565C} for the specific case of the formation of
hydrogen molecules.  Here, the fraction of binding sites with deep
potential wells totally changes the character of the problem since
numerous H atoms are present over wide temperature ranges on each
grain.  Whether or not an approach of this complexity can be extended
to large models is unclear. A simpler approach to different binding
sites is the physisorption-chemisorption model of
\cite{2004ApJ...604..222C}, where only two types of binding sites are
considered for H atoms, shallow and deep.  In this approximation, rate
equations can be used to a reasonable degree of accuracy.

At later stages, the surface chemistry includes deuteration via atomic
deuterium, which is produced in the gas by dissociative recombination
of deuterated ions; viz.,
${\rm      D_{3}^{+}  +  e^{-}  \longrightarrow  D + D + D},$
and can lead in very dense regions to an abundance of atomic deuterium
nearly equal to the abundance of atomic hydrogen
\citep{2003ApJ...591L..41R,2004A&A...424..905R}.  The H and D atoms
land on grains and, because of their rapid diffusion and high surface
reactivity, both hydrogenate and deuterate heavy atoms and reactive
molecules to form more saturated forms via sequential reactions.  For
example, the formation of water ice can occur by the addition of two
hydrogen atoms to an oxygen atom that lands on a grain surface:
${\rm O  +  H  \longrightarrow OH;  OH  +  H  \longrightarrow H_{2}O}$.
or can start from OH, which can be formed on grain surfaces by successive
reactions involving O$_2$ and O$_3$ \citep{1982A&A...114..245T}. Note
that even if the gaseous O$_2$ is low, an appreciable abundance of
O$_2$ can built up on the surface through reactions of O plus O, or
O$_2$ can be formed in the reaction of O$_3$ with H.
There is even laboratory evidence that hydrogenation of CO into
formaldehyde (H$_{2}$CO) and methanol (CH$_{3}$OH) occurs
\citep{2003ApJ...588L.121W,2004ApJ...614.1124H}: ${\rm CO \rightarrow
HCO \rightarrow H_{2}CO \rightarrow H_{3}CO \rightarrow CH_{3}OH }$ ,
although two of the hydrogenation reactions possess small activation
energy barriers. Following a similar path, not only H$_2$CO and
CH$_3$OH are formed but also their deuterated isotopologues
\citep{1997ApJ...482L.203C}.  It is interesting to note that there is
currently no viable formation mechanism for methanol in the gas phase,
so that all methanol detected in the gas has its genesis on grain
surfaces.  This statement is true even for the small amount of
methanol detected in cold sources. Here surface formation is followed
by inefficient desorption.
The results of a number of models -- which differ in the
sophistication of the gas-grain interaction, their assumptions on 
initial conditions including gas-phase abundances, and the
chemical routes included -- have been reported in the literature
\citep{1982A&A...114..245T,1992ApJS...82..167H,2003ApJ...593..906A,2003MNRAS.340..983S,2005ApJ...620..330A}.
Deuteration on cold surfaces was first explored by
\cite{1983A&A...119..177T} and later studied by
\cite{1989MNRAS.237..661B}, \cite{1997ApJ...482L.203C},
\cite{2002P&SS...50.1257C}, and \cite{2003MNRAS.340..983S} among
others.  The basic picture adopted by astrochemists is that D atoms
can deuterate atoms and radicals in a similar manner to the reactions
of H atoms.  If we, for example, consider the competing deuteration
and hydrogenation processes starting with surface CO, all of the
different isotopologues and isotopomers of methanol can be produced:
CH$_{3}$OD, CH$_{2}$DOH, CH$_{2}$DOD, CHD$_{2}$OH, CHD$_{2}$OD,
CD$_{3}$OH, and CD$_{3}$OD \citep{1997ApJ...482L.203C}.  Their
relative abundances on grain surfaces depend on the flux ratio between
H and D atoms landing on the grains.  That the deuteration is more
complex than envisioned by astrochemists, however, is shown by some
recent experimental evidence provided by \cite{2005ApJ...624L..29N}
where it seems that even normal methanol can be deuterated by
reactions with atomic D on grain surfaces.

All model calculations confirm that ices are produced by surface
reactions, and that these ices are, at least on average, dominated by
water, as observed.  The large amount of CO ice detected by observers
is probably mainly the result of accretion from the gas, where this
molecule is produced copiously.  The production of surface methanol
occurs once there is sufficient CO to be hydrogenated.  The production
of the other very abundant surface species -- carbon dioxide
(CO$_{2}$) -- is not well understood.  
%Thermal diffusive chemistry
%would require the rapid diffusion of the heavy species O and CO, which
%may or may not occur sufficiently at 10 K.  Laboratory experiments
%also indicate an activation energy barrier for the reaction, although
%the experiments currently differ as to whether or not CO$_{2}$ is
%actually produced \citep{2001ApJ...555L..61R}. Another possibility
%advocated by some is a non-thermal surface chemistry activated by
%photons or energetic particles.  Indeed, a large number of papers on
%this subject have appeared in the literature, based on laboratory
%experiments with intense fluxes of radiation (see, for example,
%%\cite{2004A&A...413..209M}, \cite{2005AdSpR..36..184H}, and
%\cite{2005ApJ...634..698B}).Whether or not the high flux, short
%lifetime laboratory measurements adequately reproduce the very
%different interstellar environment, in which low fluxes and long
%lifetimes are the rule, remains to be seen.  Detailed modeling is
%clearly needed.

\subsection{Mantle evaporation and complex organic molecule
 formation}\label{sec:mantle-evap-compl}

A variety of processes can return ice mantle species to the gas phase,
even during the cold prestellar core phase. The most important one
here is cosmic ray driven desorption where an Fe-cosmic-ray ($E\sim
10-100$ MeV/nucleon) temporarily heats the grain to a high enough
temperature that some mantle species evaporate
\citep{1983A&A...123..271L,1985A&A...144..147L}. By itself, this
process will only return weakly bound species (eg., CO, N$_2$) to the
gas phase. However, this process becomes very effective, even for more
tightly bound species such as H$_2$O, if the ices contain chemical
energy in the form of stored radicals produced by UV photolysis
\citep{1982A&A...109L..12D,1986A&A...158..119D,2005pcim.book.....T}.
% For a dense cloud
%core, the FUV field produced by cosmic ray ionization and subsequent
%excitation of H$_2$ by primary and secondary electrons is sufficient
%to maintain a radical abundance of $\sim 10^{-2}$ in the ice. Cosmic
%ray hits can then drive off water and maintain an abundance of $\sim
%0.1-1\, n^{-1}$ (for $n>10^5$ cm$^{-3}$) of condensible species in the
%gas phase \citep{2004ApJ...603..159B,2005pcim.book.....T}. 
It should be emphasized that the thermal shock driven into the grain
by the penetrating cosmic ray can also ``lift'' off a small quantity
of larger species \citep{1991ApJ...379L..75J}.

As the protostellar stage commences and temperatures begin to
increase, the ``dirty'' grain mantles will be lost partially or
totally via evaporation in a rather complex process stemming from the
heterogeneous nature of the mantles
\citep{2004MNRAS.354.1141V}. Nearer to the bipolar outflows associated
with protostellar regions, shock waves can also lead to the removal of
the mantles.  In particular, heating of ice mantles will lead to
evaporation and an injection into the warm dense gas of large
quantities of grain surface produced species such as water, methanol,
and formaldehyde. Reactions among these simple molecules can then lead
to more complex species
\citep{1991ApJ...369..147M,1992ApJ...399L..71C,1993ApJ...408..548C,2003ApJ...585..355R,2004A&A...414..409N}. The
process begins with the production of H$_3^+$ via cosmic ray
bombardment of H$_2$, followed by the reaction of H$_{2}^{+}$ and
H$_{2}$.  This triatomic species, possibly through some
intermediaries, transfers its proton to methanol. It has long been believed
that the resulting protonated methanol, CH$_3$OH$_2^+$, will rapidly
undergo alkyl cation transfer reactions with other species. With
CH$_3$OH and H$_2$CO, this might lead to the formation of dimethyl
ether and methyl formate, which are abundant complex molecules in hot
corinos (Section \ref{sec:hot-corinos}).  However, the above reactions
advocated to produce methyl formate do not occur, based on laboratory
experiments \citep{2004ApJ...611..605H}.  A more successful synthesis
seems to be the one leading to dimethyl ether:
  \begin{equation}
  {\rm CH_{3}OH + CH_{3}OH_{2}^{+}  \longrightarrow  CH_{3}OHCH_{3}^{+} + H_{2}O},
  \end{equation}
  \begin{equation}
  {\rm CH_{3}OHCH_{3}^{+}  + e^{-} \longrightarrow CH_{3}OCH_{3} + H},
  \end{equation}
but it is to be noted that experimental measurements of the neutral
products of dissociative recombination reactions show that two-body
products such as shown here are often minor channels.  Eventually,
these complex species are destroyed on a timescale of $10^{4-5}$ yr,
depending on the temperature and density of the gas, to reform
CO. Indeed, if the chemistry were understood well enough, abundances
of these species have the potential to be a chemical ``clock'', timing the
formation of the central object, which has heated up and evaporated the
ices. Other determinants of age are abundance ratios of sulfur-bearing
species, although the use of these indicators requires sophisticated
modeling of both the chemical and the physical structure of the
studied object \citep{2004A&A...422..159W,2005A&A...437..149W}.  The
need for such a sophisticated approach arises because the abundance
ratios of sulfur-bearing species, used as clocks, turn out to also be
dependent on initial abundances following evaporation as well as on
physical conditions.

Deuterium fractionation in the ices arises mainly from reactions
involving deuterium atoms accreted from the highly fractionated gas.
Such fractionation occurs in the abundant surface species (H$_2$O,
H$_2$CO, CH$_3$OH, NH$_3$, CH$_4$) acting as signposts of a cold
prestellar-core history in which gaseous D atoms are abundant.  When
these species are returned to the gas, the fractionation becomes more
apparent.  Although it is sometimes difficult to disentangle purely
gas-phase fractionation from surface fractionation followed by
evaporation, fractionation in gaseous molecules formed mainly on
surfaces (e.g., methanol) is due at least initially to surface
reactions.  This fractionation will also be passed on to daughter
products such as dimethyl ether and methyl formate via gas-phase
processes. Thus, the deuteratium fractionation can be used as a tracer
of the chemical routes involved in the formation of complex molecules
in hot cores. Indeed, the fractionation pattern of dimethyl ether and
methyl formate should reflect directly that of methanol and
formaldehyde. Eventually, reactions in the warm dense gas of the hot
core will reset this deuterium fractionation to that appropriate for
the gas temperature \citep{1997ApJ...482L.203C}. This chemical kinetic
evolution may run differently for different species or even for
different chemical groups in one species. Specifically, for the
fractionation pattern of methanol, it has been suggested that
protonation of methanol isotopomers followed by dissociative electron
recombination reforming methanol will preferentially remove the
deuteration from the OH group but not affect the CH$_3$ group
\citep{2004A&A...421.1101O}. The analysis may be incorrect, however,
because the most recent storage ring experiments
\citep{2004ApJ...613.1302G} show that protonated and deuterated
methanol ions undergo dissociative recombination reactions that lead
to virtually no neutral methanol or its isotopologues at all.  If all
of the gas-phase isotopologues of methanol are depleted at the same
rate, then the abundance ratios are indeed determined by the surface
chemistry during the previous cooler eon.

%%%%%%%%%%%%%%%%%%%%%%%%%%%%%%%%%%%%%%%%%%%%%%%%%%%%%%%%%
\section{CONCLUSIONS: CLOSED AND OPEN QUESTIONS}\label{sec:conclusions}
%\bigskip

In this contribution, we have summarized new insights into the physical
and chemical composition and evolution of Prestellar Cores and Class 0
low-mass Protostars. In particular,

1. On the physical structure, Prestellar Cores can be described by
 the density distribution of a Bonnor-Ebert sphere, whereas Class 0
 sources are consistent with the Shu ``inside-out'' recipe.

2. At the center of Prestellar Core condensations, densities are so
  high and temperatures so low that possibly all heavy-element-bearing
  molecules freeze out onto the dust grains, forming mantles of ices.
  We discussed the possibility that small grains coagulate into larger
  grains during this phase.

3. A similar molecular depletion is observed in large regions of
  the outer envelopes of Class 0 sources. These regions are
  indistinguishable from Prestellar Cores on this basis, arguing that
  the latter are precursors of Class 0 sources.

4. The molecular depletion is accompanied by a dramatic enhancement
  of the molecular deuteration, observed in both Prestellar Cores and
  Class 0 sources. In the innermost regions of Prestellar Cores, the
  only observable molecules may be H$_2$D$^+$ and HD$_2^+$, which are
  likely as abundant as H$_3^+$ or even more so.

5. In Class 0 sources, many multiply deuterated molecules have been
  observed. Even triply deuterated molecules have been detected, with
  abundances enhanced by up to 13 orders of magnitude with respect to
  the D/H elemental abundance.

6. Triggered by the observations of multiply deuterated molecules
  in Prestellar Cores and Class 0 sources, a new class of models for
  deuteration has been developed in the last few years. These models
  predict that the low temperatures coupled with the disappearance of
  heavy-element-bearing molecules from the gas phase cause the
  formation of abundant H$_2$D$^+$, HD$_2^+$ and D$_3^+$, which are
  extremely efficient in passing their deuterium atoms on to other
  molecules. In extreme conditions, D$_3^+$ is predicted to be the
  most abundant molecular ion in the gas, surpassing by orders of
  magnitude the number density of H$_3^+$.

7. During the cold phases of low-mass star formation, grain mantles
  are formed, consisting mainly of water ice, but with many other
  molecules in quantities that can be important.  Extreme molecular
  deuteration is a clear hallmark of this phase and is recorded in the
  mantles. Hydrogenated molecules, like formaldehyde and methanol, are
  also believed to be formed on the grain surfaces during the cold
  pre-collapse phase.

8. Class 0 envelopes consist of two chemically distinct regions: the
  outer envelopes and the inner regions, called hot corinos. The
  border between the two is where the dust temperature reaches the
  sublimation temperature of the grain ices. This has a major impact
  on the gas chemistry. In the hot corinos, the chemistry is dominated
  by the evaporation of the mantles, built up during the pre-collapse
  phase. Once the mantle components are in the gas phase, they undergo
  successive reactions leading to the formation of many complex
  organic molecules, similar to those observed in the hot cores.\\
%\end{itemize}

Although important progress has been achieved in these last few years,
it has also been accompanied by the rise of numerous new
questions. For example, while molecular depletion is now an accepted
``fact'', it is still not totally clear whether in the innermost
regions of the Prestellar Cores the density increases so much that not
even a trace of heavy-element species remains in the gas phase. Also,
it still remains unclear when and exactly how  the collapse starts.  It
is clear now that observations of heavy-element-bearing molecules may
not be the best way to proceed. In contrast, as we now realize, the
ground-state transition of ortho-H$_2$D$^+$ (and, to a lesser extent,
para-HD$_2^+$) provides a new and exciting probe with which to search
for the long-sought collapse in the innermost regions of Prestellar
Cores \citep{2005A&A...439..195V}.  However, in order to fully exploit
this new diagnostic, we have to first fully understand the ortho-para
ratio of H$_2$D$^+$ \citep{2006astro.ph..1429F}, which is not measured
at all so far.  It is also clear that models for the Prestellar Core
phase have improved dramatically with the inclusion of the extreme
depletions of the gas phase and the formation of the multiply
deuterated forms of H$_3^+$. However, these models still struggle to
reproduce the extreme observed deuteration, especially that of
formaldehyde and methanol. While this is likely due to grain
surface chemistry, it remains the case that, unless there is an
enhancement in the reactivity of D over H, the gas landing on the
grain surfaces at the moment of the mantle formation must have an
atomic D/H ratio higher than accounted for by the current gas phase
chemistry models. We also have little knowledge as to whether the
prediction that D$_3^+$ is the most abundant molecular ion in extreme
conditions is correct. The coming years will likely see many of these
questions answered, with the advent of the satellite Herschel and,
foremost, the interferometer ALMA.

If many questions remain about the cold pre-collapse phase, even more
questions are unanswered for the protostellar phase, and many have
likely not even been asked. For example, the very nature of the hot
corinos is still very much debated. Several authors question how much
of the observed warm gas, with complex organic molecules, is due to
the interaction of an outflow with the inner envelope
\citep{2005ApJ...632..371C}. Another possibility invoked is that this
warm gas enriched with complex organic molecules resides in a
disk-like atmosphere \citep{2005ApJ...632..973J}. The reality is
indeed that very few observations have resolved hot corinos; actually,
 in just one source has this so far occurred
\citep{2004ApJ...616L..27K,2004ApJ...617L..69B}. Until more sensitive
observations are available with SMA and ALMA, the question will remain
open for debate. Another open question regards the chemistry leading
to the observed molecular complexity in hot corinos. No clear
observations are able so far to distinguish the roles of gas phase and
grain surface chemistry in the formation of formic acid or methyl
formate, just to mention two molecules frequently detected. The answer
to this question will likely need not only more and more sensitive
observations towards the hot corinos, but also laboratory experiments
coupled with modeling in order to elucidate the role of the grains in
the story. Last, but not least, we suspect that we are just viewing
the tip of the iceberg: what is still hidden from our eyes?  What is
the ultimate molecular complexity reached during the hot corino phase?
Do biotic molecules form? Do they play a role in the trigger and/or
diffusion of life in the Universe (see the chapter by
{\it Gaidos \& Selsis})?

Finally, many more questions exist regarding the fate of the molecules
formed through the pre-stellar and protostellar phases. It is likely
that some condense again onto the grain mantles during the phase of
the proto-planetary disk, at least in the outer zones of the disk that
are shielded by FUV photons (see the chapters by {\it
Bergin et al.}, {\it Dutrey et al.} and {\it Dullemond et al.}). The
same grains are likely to coagulate into larger aggregates, which are
the bricks from which the planets eventually form (see the chapters by {\it Dominik et al.} and {\it Wadhwa et al.}). How much of the
mantle formed in the Prestellar Core phase is preserved? Do the
molecules, and particularly the complex organic molecules formed
during the hot corino phase, freeze-out onto the new mantles without
other alterations? Or what kind of alterations do they undergo? How
pristine is the material forming the comets, asteroids and planets?

To reconstruct the full story, we need many more pieces
of the puzzle.  One strong motivation for pursuing the story of
protoplanetary disks is that their aftermath may have been part of the
history of our own Solar System. The water of the oceans is an
emblematic example. The ratio between HDO and H$_2$O (known as SMOW)
is $\sim{10}^{-4}$ \citep{dewiietal1980}, fully ten times larger than
the D/H elemental abundance of the Solar Nebula \citep{geiss1993}. The
origin of the oceans has been long debated and new theories keep
coming. The deuterium enhancement is in line with the theory that much
of the ocean water comes from the bombardment of the early Earth by
comets \citep{1995Icar..116..215O}, for comets show a similarly
enhanced HDO/H$_2$O ratio
\citep{1998Sci...279..842M,1998Icar..133..147B}. Other possibilities
include water brought by smaller bodies of the Solar System
(meteorites and asteroids)
\citep{2004Icar..168....1R,2005Natur.435..466G}. These bodies, as well
as comets, have the imprint of the pre-collapse/protostellar phase in
their ices (see the chapters by {\it Jewitt et al., Wooden
et al.} and {\it Alexander et al.}).  Regarding deuterium fractionation,
the question remains as to what extent the enhanced HDO/H$_2$O ratio
is a legacy from the protostellar/proto-planetary phase and to what
extent it has been modified by subsequent chemistry.  And in addition
to water, what other molecules were inherited by the early Earth and
how did they arrive? As with other molecular indicators, how much of
the terrestrial HDO/H$_2$O ratio is a legacy depends on how it evolved
during the various phases of the earth's formation. So far, we have
constraints on the evolution of deuterium fractionation from the
pre-collapse, protostellar, and proto-planetary phases for only two
molecules, HCO$^+$ and H$_2$O, and in the outer zones of the disk
only, which may not necessarily be connected with cometary
formation. The case of DCO$^+$/HCO$^+$, representing deuteration in
the gas phase, seems to keep a rather constant value, around 10\%
\citep{2003A&A...400L...1V}. The HDO/H$_2$O ratio, very likely a
product of the grain mantles, seems to remain quite similar too, at a
level of $\sim$1\%, with just a relatively small decrease in the
proto-planetary phase
\citep{2005ApJ...631L..81C,2005ApJ...635L..85D}. Evidently, this
information alone is far too little to resolve the problem of the
oceanic origin, but shows how important the study of chemistry in the
prestellar and protostellar phases of star formation is and how much
it will eventually aid our understanding of later phases up through
the present eon. The coming decade will likely see many of these
questions answered with the advent of the satellite Herschel and the
sub-millimeter interferometer ALMA.

\smallskip
\textbf{ Acknowledgments.} We wish to thank the referee for valuable
comments on the manuscript. This study was supported in part by the
European Community's Human Potential Programme under contract MCRTN
512302, Molecular Universe, the French ``Projet National PCMI'', the
NASA Planetary Geology and Geophysics Program under grant NAG 5-10201.

\bigskip
%\newpage
%\centerline\textbf{REFERENCES}
%\bigskip

\parskip=0pt
{\small
\baselineskip=11pt

%\bibliographystyle{ppv}
%\bibliography{/Users/cecilia/common/ceccarellic}

}% end of small

\end{document}